# Fully Homomorphic Encryption Encapsulated Difference Expansion for Reversible Data hiding in Encrypted Domain

Yan Ke, Min-qing Zhang, Jia Liu, Ting-ting Su, Xiao-yuan Yang

*Abstract*—This paper proposes a fully homomorphic encryption encapsulated difference expansion (FHEE-DE) scheme for reversible data hiding in encrypted domain (RDH-ED). In the proposed scheme, we use key-switching and bootstrapping techniques to control the ciphertext extension and decryption failure. To realize the data extraction directly from the encrypted domain without the private key, a key-switching based least-significant-bit (KS-LSB) data hiding method has been designed. In application, the user first encrypts the plaintext and uploads ciphertext to the server. Then the server performs data hiding by FHEE-DE and KS-LSB to obtain the marked ciphertext. Additional data can be extracted directly from the marked ciphertext by the server without the private key. The user can decrypt the marked ciphertext to obtain the marked plaintext. Then additional data or plaintext can be obtained from the marked plaintext by using the standard DE extraction or recovery. A fidelity constraint of DE is introduced to reduce the distortion of the marked plaintext. FHEE-DE enables the server to implement FHEE-DE recovery or extraction on the marked ciphertext, which returns the ciphertext of original plaintext or additional data to the user. In addition, we simplified the homomorphic operations of the proposed universal FHEE-DE to obtain an efficient version. The Experimental results demonstrate that the embedding capacity, fidelity, and reversibility of the proposed scheme are superior to existing RDH-ED methods, and fully separability is achieved without reducing the security of encryption.

## I. INTRODUCTION

REVERSIBLE data hiding in encrypted domain (RDH-ED) is an information hiding technique that aims to not only accurately embed and extract the additional messages in the ciphertext, but also restore the original plaintext losslessly [1][2]. RDH-ED is useful in some distortion intolerable applications, such as ciphertext management or retrieval in the cloud, ciphertext annotation for medical or military use. With the increasing demand for information security and the development of the encrypted signal processing techniques, RDH-ED has been an issue of great attention in the field of privacy protection and ciphertext processing.

From the viewpoint of the cryptosystem that RDH-ED methods are based on, existing RDH-ED methods could be classified into two categories: Symmetric encryption based RDH-ED [1], [3]- [14], and public key encryption based RDH-ED [20]-[29]. Symmetric cryptography that has been introduced into RDH-ED includes stream encryption [1], [3]-[6], [13]- [13], advanced encryption standard (AES) [7], [8], and RC4 encryption [9].

According to the methods of utilizing the redundancy in the cover for data hiding, symmetric encryption based RDH-ED methods were classified into two categories [1][2]: "vacating room before encryption (VRBE)" [1][7][8][10][13] and "vacating room after encryption (VRAE)"[3]-[6]. The room, namely the redundancy in the cover, is vacated for reversible data hiding. The first RDH-ED method was proposed by Zhang for encrypted images [3], and then [4]-[5] enhanced its capacity. Qian *et al.* proposed a similar method to embed data in an encrypted JPEG bit stream [6]. AES was introduced in [7] to encrypt the cover image. Each block containing *n* pixels could carry one bit data. The embedding rate (ER) is $1/n$ bits per pixel (bpp). Then difference prediction was introduced before encryption in [8], and AES was used to encrypt pixels except the embedding ones, thus resulting in a better embedding capacity (EC) and reversibility. However, it needed decryption first before data extraction in the above RDH-ED methods, which restricted the practicability in practical applications. The separable RDH-ED was proposed in [11][12]. Separability has been so far an important attribute of practicality for current RDH-ED.

The redundancy introduced by VABE or VARE is independent from the encryption, resulting in the mutual restriction between decryption distortion and the embedding capacity, which is a major obstacle to the realization of separability and a high EC. There existed two main solutions proposed: one is to improve the quality of redundancy introduced before encryption. For example in [13], a separable high embedding algorithm was proposed by making full use of prediction error introduced before encryption. Second, the correlation of the plaintext is preserved in the ciphertext, so that RDH in spatial domain, such as difference expansion technique (DE) [15], histogram shifting technique (HS) [16]-[18], could be implemented in the encrypted domain. For example in [14],

This work was supported in part by National Key R&D Program of China under Grant 2017YFB0802000 and the National Natural Science Foundation of China under Grant 61379152, 61872384, and 61403417.

Yan Ke, Minqing Zhang, Jia Liu, Tingting Su, and Xiaoyuan Yang are with the Key Laboratory of Network and Information Security Under the Chinese People Armed Police Force (PAP), College of Cryptography Engineering in Engineering University of PAP, Xi'an, 710086, China (e-mail: 15114873390@163.com; api_zmq@126.com; twinly77@gmail.com; suting0518@163.com; yxyangyxyang@163.com).

a new framework of RDH-ED was proposed, in which a specific stream cipher was used to preserve the correlation between the neighboring pixels. The above mentioned symmetric encryption based algorithms are fast and efficient in practice, which has significant research value and technological potential in the future.

However, there are also technical defects in symmetric encryption based RDH-ED. The correlation of plaintext would be destroyed because of the *confusion* and *diffusion* principles of symmetric encryption. To achieve reversible data hiding, it usually needs to introduce embedding redundancy. While it is difficult to vacate room after encryption, the current attention focuses more on the VEBE methods [13], by which more computational expense is introduced into the client end for data hiding. The preprocessing in the plaintext is similar to data compression, and the compression capability determines the performance of RDH-ED. As for the methods of preserving plaintext correlation by a specific encryption [14], it currently mainly relies on reusing the same random sequence to encrypt a specific pixel block. It could provide certain security guarantees, but key reusing would weaken the encryption intensity of the symmetric encryption in theory. The more correlation among ciphertext is remained, the more the encryption intensity is reduced. The RC4 encryption was declared breached in 2013 [19], RDH-ED based on early RC4 has certain limitations in future security applications. In addition, symmetric encryption requires a geometrically increasing amount of encryption keys with the number of communication participants. The local key storage cost is high for each user.

Compared with symmetric encryption, public key encryption has some advantages for RDH-ED, which is worthy of our attention: first, public key encryption requires a linear increasing amount of key usage in the communication network. The local key storage cost is only the private key of the user's own, while all the public keys are publicly released. It has been widely used in electronic finance and network communication protocols, which provides application prospects for RDH-ED. Second, public key encryption introduces ciphertext extension, namely, the redundancy from the ciphertext itself. Through a certain embedding strategy [28], we could select embedding positions and improve EC effectively. Third, flexible cryptosystems of the public key encryption, especially the homomorphic encryption, provide reliable technical supports for RDH-ED. However, there are still technical limitations and application dilemmas in public key based RDH-ED. We shall discuss those in Section II. This paper focuses on the current state of public key based RDH-ED, aiming at making full use of LWE-based fully homomorphic encryption (FHE) technique to implement DE encapsulation. A novel RDH-ED method is proposed, which is superior to the current public key based RDH-ED in practicality, security and reversibility.

The rest of this paper is organized as follows. The following section introduces the art of state about public encryption based RDH-ED and analyzes the potential of DE for RDH-ED. Section III introduces the techniques of FHE, key-switching, and bootstrapping. Section IV describes the detailed processes of the proposed full homomorphic encryption encapsulated difference expansion. In Section V, the three judging standards of RDH-ED, including *correctness*, *security* and *efficiency*, are discussed theoretically and verified with experimental results. Finally, Section VI summarizes the paper and discusses future investigations.

## II. RELATED WORK

### A. Public Encryption Based RDH-ED

Currently, researches of public key encryption based RDH-ED are mainly based on Paillier encryption [20]-[26] and learning with Error (LWE) encryption [27]-[29]. Probabilistic and homomorphic properties of the above cryptography allow the third party, *i.e.*, the cloud servers, to conduct operations directly on ciphertext without knowing the private key, which shows potential for more flexible realizations of RDH-ED.

The first Paillier encryption based RDH-ED was proposed by Chen *et al.* [20]. Shiu *et al.* [21] and Wu *et al.* [22] improved the EC of [20] by solving the pixel overflow problem. Those algorithms were VRBE methods. Li *et al.* in [25] proposed a VRAE method with a considerable EC by utilizing the homomorphic addition property of Paillier encryption and HS technique. The above algorithms were all inseparable. Data extraction was implemented only in the plaintext domain. It was a crucial bottleneck of public key encryption based RDH-ED to realize data extraction directly from the encrypted domain. Wu *et al.* proposed two RDH-ED algorithms for the encrypted images in [24]: a high-capacity algorithm based on Paillier cryptosystem was presented for data extraction after image decryption. The other one could operate data extraction in the encryption domain. Zhang *et al.* [23] proposed a combined scheme consisting of a lossless scheme and a reversible scheme to realize separability. Data was extracted from the encrypted domain in the lossless scheme and from the plaintext domain in the reversible scheme. In [26], Xiang embedded the ciphertext of additional data into the LSBs of the encrypted pixels by employing homomorphic multiplication. Only the ciphertext of additional data could be obtained during extraction directly from ciphertext. To distinguish the corresponding plaintext of the ciphertext of additional data without the private key, a one-to-one mapping table from ciphertext to plaintext was introduced while the ciphertext of additional data for embedding was not from encryption but from the mapping table. However, the exposure and accumulation of a large number of the mapping tables to an untrusted third party might increase the risk of cryptanalysis in theory, while the Paillier algorithms cannot resist *adaptive chosen ciphertext attack* (ACCA or CCA2) [30].

LWE based RDH-ED was first proposed in [26] by quantifying the LWE encrypted domain and recoding the redundancy from ciphertext. Ke *et al.* fixed the parameters for LWE encryption and proposed a multilevel RDH-ED with a flexible applicability and high EC in [27]. However, the data-hiding key used for extraction overlapped partly with the private key for decryption, thus resulting in limitation for embedding by a third party. In [29], separability could be achieved by preserving correlation from the plaintext in the ciphertext through a modified somewhat LWE encryption. However, the correlation among ciphertext was strong, and it was theoretically vulnerable to cryptanalysis attacks.

In summary, the public key based RDH-ED has difficulties in the implementation of separability. Usually, EC and the fidelity are inferior to existing RDH in spatial domain. Therefore, this paper proposes a novel scheme of FHE encapsulated DE (FHEE-DE) to realize data hiding in encrypted domain based on LWE. A fidelity constraint is proposed for FHEE-DE, which has distinctly enhanced the Peak Signal-Noise Ratio (PSNR) of the directly decrypted images. Modified bit-addition and bit-subtraction circuits for FHEE-DE are designed. Key-switching and bootstrapping are introduced to control the ciphertext extension and decryption failure, which provides the feasibility of introducing other existing RDH methods. To realize separability, a key-switching based least significant bit data hiding (KS-LSB) method is proposed to ensure the extraction directly from the encrypted domain without the private key. Finally, to improve the efficiency, we propose an efficient FHEE-DE besides the universal one. Experimental results demonstrated that the performances of the proposed scheme in EC, fidelity, and reversibility are superior to DE in [15] and existing public key based RDH-ED methods, and fully separability is achieved without reducing the security of LWE encryption.

*B. Difference Expansion*

The difference expansion technique is an important part of the early RDH algorithms, appearing together with the histogram shifting technique. It is characterized by specific numerical modification of specific pixels, making it reversible under the premise of fault tolerance. However, due to the image distortion, the degree of modification on pixels is limited. The constraints of selecting available pixel pairs increase with the requirements of reversibility and EC of RDH technology, which currently has restricted the development of DE algorithms. The HS based RDH and its variant algorithms have become the focus in image spatial domain. However, considering the potential of being encapsulated by FHE, DE has more advantages than HS:

1. The less computational and memory consumption. The object of DE is one specific pair of pixels while that of HS relies on the statistical characteristics of the image pixels as a whole. Therefore, the object of FHEE-DE is specific pairs of encrypted pixels while that of FHE encapsulated HS might be the plaintext statistical characteristics obtained by processing in the encrypted domain. The computational and memory consumption of calculating statistical characteristics from the ciphertext might be much larger, because the number of pixels involved in the calculation is large, and it has to store a large number of logical decisions (the results cannot be logically compared directly in the encrypted domain, and multiple possible states need to be saved at the same time).

2. The higher accuracy of cover recovery. The reversibility of DE is the deterministic reversibility of maintaining the correctness of arithmetic operations, while the reversibility of HS is the probabilistic reversibility of maintaining histogram statistical features. Therefore, DE has advantages in maintaining the accuracy of cipher recovery after homomorphic operations.

In addition, there are many interesting implementations [31][32] of DE algorithms. DE algorithms and their variants should be suitable for being introduced into RDH-ED through FHE technology.

This paper started with the earliest DE algorithm [15] and proposed a scheme of FHE encapsulated DE, aiming to provide a technical foundation for the introduction of more DE based algorithms in the later stage.

In Tian's algorithm [15], two adjacent pixels $X$ and $Y$ from an image $I$ can be used to hide one additional bit $b_s$, where $0 \leq X, Y \leq 255$ and $b_s \in \{0, 1\}$. First, the difference $h$ and average value $l$ (integer) of $X$ and $Y$ are computed as following:

$$h = X - Y \qquad (1)$$

$$l = \left\lfloor \frac{X+Y}{2} \right\rfloor \qquad (2)$$

$$X = l + \left\lfloor \frac{h+1}{2} \right\rfloor \qquad (3)$$

$$Y = l - \left\lfloor \frac{h}{2} \right\rfloor \qquad (4)$$

Assuming $X > Y$, and $\lfloor . \rfloor$ is the floor function meaning "the biggest integer less than or equal to" while $\lceil . \rceil$ is the ceiling function.

Data hiding:

$$h' = 2 \times h + b_s \qquad (5)$$

The embedded pixels $X'$ and $Y'$ can be obtained by substituting $h'$ into Eqs. (3), (4).

Extration:

$$b_s = LSB(h') \qquad (6)$$

$LSB(.)$ is to obtain the least significant bit of the input integer.

Recovery:

$$h = \left\lfloor \frac{h'}{2} \right\rfloor \qquad (7)$$

Then, the original pixels $X$ and $Y$ can be recovered by using Eqs. (3), (4).

III. PRELIMINARIES

*1) Full homomorphic encryption*

A cryptosystem that supports arbitrary computation on ciphertext is known as fully homomorphic encryption. Such a scheme enables the construction of programs for any desirable functionality, which can be run on encrypted inputs to produce an encryption of the result. Since such a program never needs to decrypt its inputs, it can be run by an untrusted party without revealing its inputs and internal state. Therefore, FHE has great practical advantages on outsourcing the private computations for reversible data hiding.

Craig Gentry, [33] using LWE in lattice based cryptography, described the first construction for a FHE scheme. Gentry's scheme supports both addition and multiplication operations on ciphertext, from which it is possible to construct circuits for

performing arbitrary complex computation. A brief review of FHE [33] is as following:

The private key is denoted as $s$, and the public key $A$ is generated by $s$ and $e$ satisfying Eq. (8), where $e$ is sampled randomly:

$$A \cdot s = 2e \qquad (8)$$

*Encryption*:
The plaintext is $m \in \{0, 1\}$. Set $\boldsymbol{m} = (m, 0, 0, \ldots, 0)$. Generate a 0-1 sequence $\boldsymbol{a}_r$ uniformly and output the ciphertext:

$$\boldsymbol{c} = \boldsymbol{m} + \boldsymbol{A}^{\mathrm{T}} \boldsymbol{a}_r \qquad (9)$$

*Decryption*:

$$\left[[\langle \boldsymbol{c}, \boldsymbol{s}\rangle]_q\right]_2 = \left[[\langle \boldsymbol{m} + \boldsymbol{A}^{\mathrm{T}} \boldsymbol{a}_r, \boldsymbol{s}\rangle]_q\right]_2 = \left[\left[\boldsymbol{m}^{\mathrm{T}} \boldsymbol{s} + (\boldsymbol{A}^{\mathrm{T}} \boldsymbol{a}_r)^{\mathrm{T}} \boldsymbol{s}\right]_q\right]_2$$
$$= \left[\left[m + \boldsymbol{a}_r^{\mathrm{T}} \boldsymbol{A}\boldsymbol{s}\right]_q\right]_2 = \left[\left[m + \boldsymbol{a}_r^{\mathrm{T}} 2\boldsymbol{e}\right]_q\right]_2 = m \qquad (10)$$

where $[.]_q$ means to perform modulo $q$. The correctness lies in that the total introduced noise could be restrained to meet:

$$\boldsymbol{a}_r^{\mathrm{T}} \boldsymbol{e} < q/4 \qquad (11)$$

To demonstrate the FHE ability, we assume the LWE ciphertext $\boldsymbol{c}_1, \boldsymbol{c}_2$ are: $\boldsymbol{c}_1 = m_1 + 2r_1 + p_1 q$, $\boldsymbol{c}_2 = m_2 + 2r_2 + p_2 q$.

*FHE. Add*:

$$\boldsymbol{c}_1 + \boldsymbol{c}_2 = (m_1 + m_2) + q(p_1 + p_2) + 2(r_1 + r_2) \qquad (12)$$

where the correctness lies in that the total introduced noise could be restrained to meet:

$$(r_1 + r_2) < q/4 \qquad (13)$$

*FHE. Multiply*:

$$\boldsymbol{c}_1 \otimes \boldsymbol{c}_2 = m_1 m_2 + 2(m_1 r_2 + m_2 r_1 + 2 r_1 r_2) + \\ q(\boldsymbol{c}_2 p_1 + \boldsymbol{c}_1 p_2 - q p_1 p_2) \qquad (14)$$

where the correctness lies in that the total introduced noise could be restrained to meet:

$$m_1 r_2 + m_2 r_1 + 2 r_1 r_2 < q/4 \qquad (15)$$

*2) Key–switching* [34]
There is data expansion in LWE encrypted ciphertext [28], but in FHE, a secondary expansion would occur when ciphertext got multiplied. The homomorphic multiplication between the ciphertext matrices returns the ciphertext tensor product, and the private key is also subjected to the tensor product operation before being used to decrypt the new ciphertext. Therefore, the amount of data will again expand geometrically.

In our scheme, the ciphertext of the pixel bits will get expanded after each multiplication. It also occurs after addition or subtraction between encrypted pixels due to the cases of bit carry or bit borrow, resulting in a large number of multiplication and exclusive or operations among ciphertext of pixel bit. If the secondary expansion cannot be eliminated or controlled, the amount of ciphertext data can produce an excessively extension that is unacceptable in practice. Key-switching can effectively eliminate the extension by replacing the extended ciphertext with new ciphertext of any shorter length without decrypting it, and ensure the new ciphertext corresponds to the same decryption as the extended ciphertext.

We use the key-switching technique to eliminate the ciphertext secondary expansion in FHEE-DE, that is, key-switching is implemented following each homomorphic operation. What is more, a key-switching based LSB data hiding method is proposed in this paper.

*3) Bootstrapping Encryption* [35]
The introduced noise, on the one hand, provides the security guarantee of the LWE algorithm. On the other hand, noise superposition will also affect the correctness of decryption. Usually, the one-way fluctuation interval of required noise cannot exceed a quarter of the encrypted domain. Homomorphic operations result in the superposition of noise as shown in Eqs. (11)(13)(15), which makes the correctness of decryption unstable. Key-switching cannot eliminate noise superposition. Therefore, the decryption overflow problem is an important issue to be considered in FHEE-DE. Usually, the simplest method is to limit the standard deviation of the sampling noise distribution, so that the noise fluctuation range is very small and the overflow will not occur after several times of superposition. Although efficient, it is not enough to support many or even theoretically infinite holomorphic operations. If the number of homomorphic operations is limited, it limits the realization of homomorphic encapsulation of DE, and it is not conducive to further extending to other variants of DE.

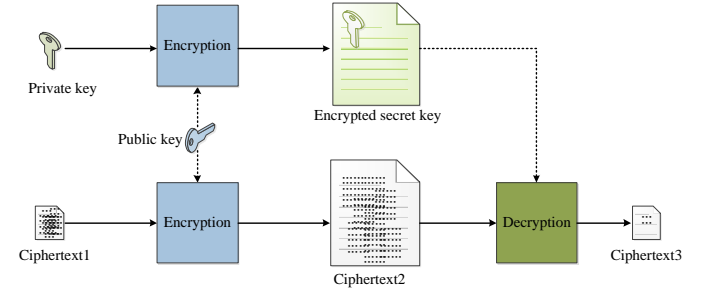

Fig. 1. The sketch of the bootstrapping process.

In this paper, we use bootstrapping to reduce the superposed noise. The sketch of bootstrapping is as shown in Fig. 1: ciphertext1 carries superposed noise. A public key is used to encrypt ciphertext1 and the private key simultaneously. The encrypted private key is then used to decrypt the ciphertext2 (the encrypted ciphertext1) to obtain ciphertext3. The decryption process eliminates the noise in ciphertext1, but retains the noise from the bootstrapping encryption in ciphertext 3. Therefore, bootstrapping does not change the data length of ciphertext or the key, but it can restore the total noise amount after several times of superposition into the noise amount introduced by only once bootstrapping encryption (for more details in [35]).

Key-switching and bootstrapping are used to ensure FHEE-DE based RDH-ED has good practicability and scalability. The use and parameter requirements of the algorithm will be introduced in detail in the following sections, and some adaptive modifications will also be made to meet the specific requirements of reversibility. The main drawback of the two techniques is the large number of public keys required. Fortunately, users only need to store the private key for decryption. The public keys are all released publicly on the internet or the cloud instead of stored locally after their generation. The storage problem caused by the large amount of public keys can be ignored in practice, which is also the advantage of public key cryptosystem in application. Besides, the operation of generating public keys is not strictly in series with other processes, so parallel optimization techniques can be used to improve the efficiency. The parallel optimization is not the consideration of this paper.

## IV. FULL HOMOMORPHIC ENCRYPTION ENCAPSULATED DIFFERENCE EXPANSION

### A. Framework of FHEE-DE

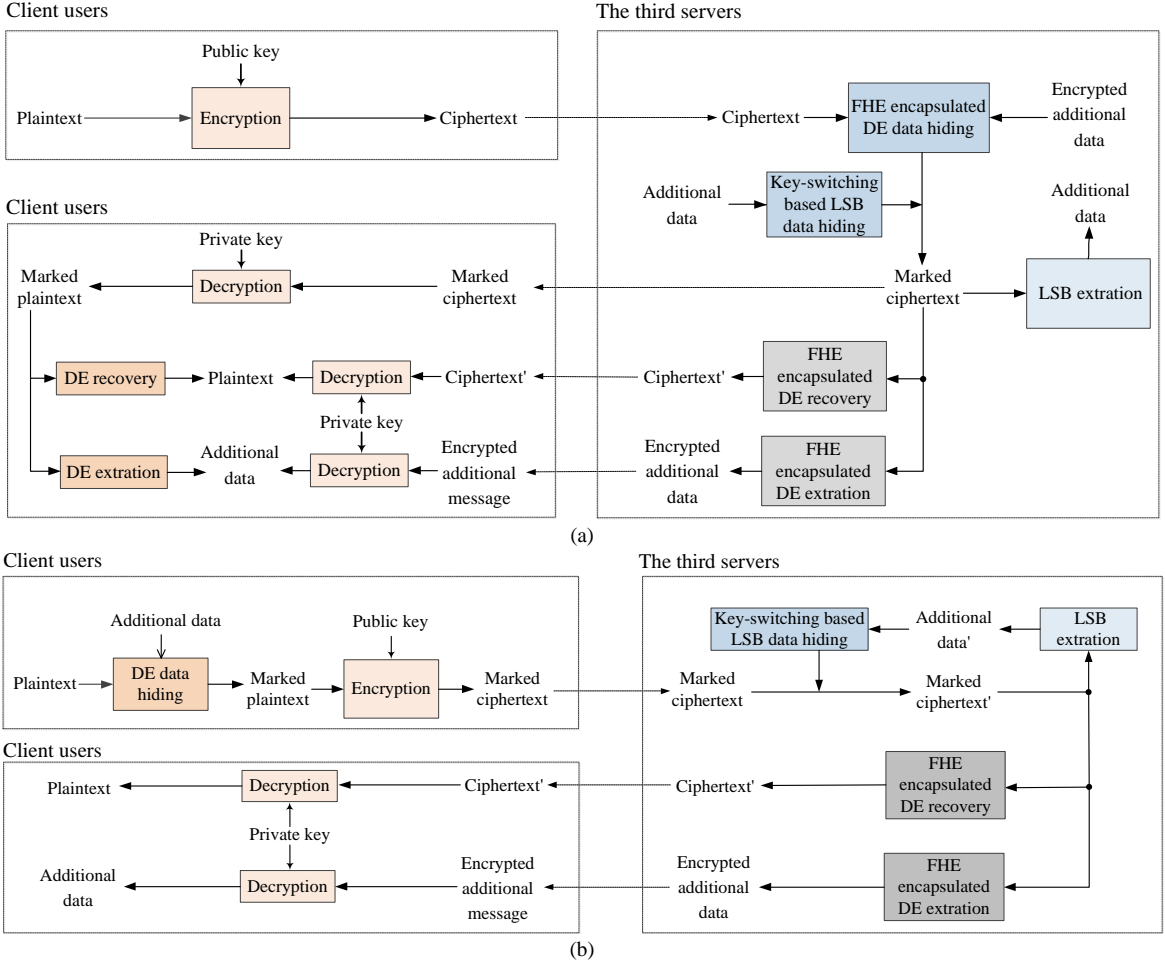

Fig. 2. The two schematic flows of the applications of FHEE-DE: (a) Data hiding in the third server side; (b) Data hiding in both the client and server side.

As shown in Fig. 2, the application framework of FHEE-DE consists of two types: *(a)* Data hiding in the third server side. *(b)* Data hiding in both the client and server side. We take *type a* as an example to explain the application: the user encrypts the plaintext and uploads its ciphertext to the server. The server performs *FHEE-DE data hiding* and *KS-LSB data hiding* to obtain the marked ciphertext. FHEE-DE data hiding ensures that the decryption would contain additional data. KS-LSB data hiding enables the server to extract additional data directly from the marked ciphertext. For the marked ciphertext, there are four cases: *a)* the client obtains and decrypts it directly to obtain the marked plaintext. *DE extraction or recovery* can be implemented to obtain additional data or the plaintext losslessly. *b)* Additional data can be extracted directly by the server without using the private key, which enables the (trusted or untrusted) third party to manage ciphertext flexibly under the premise of keeping the plaintext secret. *c)* The third server returns new ciphertext through *FHEE-DE recovery*. The client user can obtain the plaintext by decrypting the new ciphertext. *d)* FHEE-DE extraction returns the encrypted additional data. The client user can obtain the additional data by decrypting the encrypted additional data.

### B. Universal FHEE-DE

#### 1) Preprocessing of DE

*a) Overflow/Underflow and Fidelity Constraints*

The plaintext is a 512×512 image $I$. $I$ is divided into non-overlapping pixel pairs. Each pair consists of two adjacent pixels. Next, we take one pair of pixels, denoted as $(X, Y)$, as an example to introduce our scheme, where $0 \leq X, Y \leq 255$. The additional bit is $b_s \in \{0, 1\}$. As grayscale values are bounded in [0, 255], we have $h$ and $l$ according to Eqs. (1), (2) and then:

$$0 \leq l + \left\lfloor \frac{h+1}{2} \right\rfloor \leq 255 \tag{16}$$

$$0 \leq l - \left\lfloor \frac{h}{2} \right\rfloor \leq 255 \tag{17}$$

To avoid overflow or underflow problems after data hiding, we use a map matrix $\boldsymbol{M}_{\text{ava}} \in \{0,1\}^{256 \times 256}$ to indicate available pixel pairs. Value "1" indicates the bigger pixel within an available adjacent pixel pair for DE data hiding. The difference $h$ of an available pair should satisfy the following constraints [15]:

$$|h| \leq \min(2(255-l), 2l+1) \tag{18}$$
$$|2h+b_s| \leq \min(2(255-l), 2l+1) \tag{19}$$
for $b= 0$ or $1$.

When performing FHEE-DE, we add an extra fidelity constraint: the available pixel pairs are preferentially selected with a smaller pixel difference. The fidelity parameter $h_{\text{fid}}$ is introduced here as another constraint:
$$h \leq h_{\text{fid}} \tag{20}$$
$M_{\text{ava}}$ would be lossless compressed as side information of the ciphertext to superimpose on the host signal.

*b) Parameters Setting and Function Definition*

The cryptosystem is parameterized by the integers: $n$ (the length of the private key), $q \in (n^2, 2n^2)$ (the modulus), $d \geq (1+\varepsilon)(1+n)\log_2 q$ (the dimension of the public key space), $\varepsilon > 0$. If $q$ is a prime, all the operations in the cryptosystem are performed modulo $q$ in $\mathbb{Z}_q$, $\beta = \lceil \log_2 q \rceil$. We denote the noise probability distribution on $\mathbb{Z}_q$ as $\chi$, $\chi = \overline{\Psi}_{\alpha q}$, where the discrete Gaussian distribution $\overline{\Psi}_{\alpha q} = \{\lceil qx \rfloor \bmod q \mid x \sim N(0, \alpha^2)\}$, and $\lceil qx \rfloor$ denotes rounding $qx$ to the nearest integer [28].

*Definition 1*: The private key generating function:
$$s = SKGen_{n, q}(.) \tag{21}$$
which returns the private key $s \in Z_q^n$: $s = (1, t)$, where $t \in Z_q^{n-1}$ is sampled from the distribution $\chi$.

*Definition 2*: The public key generating function:
$$A = PKGen_{(d, n), q}(s) \tag{22}$$
in which a matrix $W \in Z_q^{d \times (n-1)}$ is first generated uniformly and a vector $e \in Z_q^d$ is sampled from the distribution $\chi$, then the vector $b \in Z_q^d$ is obtained:
$$b = Wt + 2e \tag{23}$$
the $n$-column matrix $A \in Z^{d \times n}$ is consisting of $b$ followed by $-W$, $A = (b, -W)$. $A$ is returned as the public key.

*Remark*: Observe that $A s = 2e$ for Eq. (8).

*Definition 3*: The encrypting function:
$$c = Enc_A(m) \tag{24}$$
which returns a vector $c$ as the ciphertext of one bit plaintext $m \in \{0,1\}$ with the public key $A$: Set $m = (m, 0, 0, \ldots, 0) \in Z_2^n$. Generate a random vector $a_r \in Z_2^d$ uniformly and output $c$:
$$c = m + A^T a_r \tag{25}$$

*Definition 4[34]*: The function $BitDe(x), x \in Z_q^n$, decomposes $x$ into its bit representation. Namely, it outputs $(u_1, u_2, u_3, \ldots, u_\beta,) \in Z_q^{n\beta}$, $x = \sum_{j=0}^{\beta-1} 2^j \cdot u_j$, $u_j \in Z_2^n$.

*Definition 5*: The decrypting function:
$$m = Dec_s(c) = \left[\left[\langle c, s \rangle\right]_q\right]_2 \tag{26}$$
which returns the plaintext bit $m \in \{0,1\}$ with the private key $s$.

If the inputs of the decryption function are in binary form, we could regard such a function as a decryption circuit, denoted as $Dec^*_S(C)$, $C = BitDe(c)$, $S = BitDe(s)$.

*Definition 6[34]*: The function $Powersof(x), x \in Z_q^n$, outputs the vector $(x, 2x, 2^2 x, \ldots, 2^{\beta-1} x,) \in Z_q^{n \cdot \beta}$.

Next, we will give the procedure of key-switching, which takes a ciphertext $c_1$ under $s_1$ and outputs new ciphertext $c_2$ that encrypts the same plaintext under the private key $s_2$.

*Definition 7[34]*: The switching key generating function:
$$B = SwitchKGen(s_1, s_2) \tag{27}$$
where $s_1 \in Z_q^{n1}$, $s_2 \in Z_q^{n2}$. $A_{\text{temp}} = PKGen_{(n1 \cdot \beta, n2), q}(s_2)$. The matrix $B \in Z_q^{(n1 \cdot \beta) \times n2}$ can be obtained by adding $Powersof(s_1)$ to $A_{\text{temp}}$'s first column.

Ciphertext $c_2$ can be obtained by using the switching key:
$$c_2 = BitDe(c_1)^T \cdot B \tag{28}$$

The secondary data expansion of the ciphertext is resulted from the homomorphic multiplication. Namely, we need to operate key-switching after each tensor product of ciphertext. Specifically, we would transform $c \otimes c$ under $s \otimes s$ into $c$ under $s$. Therefore, the switching keys for eliminating the secondary data expansion should be:
$$B = SwitchKGen(s \otimes s, s) \tag{29}$$
where $s \otimes s \in Z_q^{n \cdot n}$, $s \in Z_q^n$.

*c) Key Distribution*

In our scheme, there is a *key-switching based LSB data hiding* method proposed to ensure that the servers could directly extract additional data from ciphertext without using the private key. We generate a pseudo-random binary sequence $k$ for the servers to randomly scramble the additional data before KS-LSB data hiding. The switching key for KS-LSB data hiding is:
$$B_{\text{LSB}} = SwitchKGen(s, s) \tag{30}$$
where $s \in Z_q^n$. All different keys are distributed as shown in Table I:

TABLE I
KEY DISTRIBUTION

| Classification | Denotation | Function | Owner |
|---|---|---|---|
| Private key | $s$ | 1. Public key and switching key generation. 2. Plaintext and additional data decryption. | Client user |
| Public key | $A$ | 1. Data encryption. 2. Bootstrapping. | Servers with open access |
| Switching key | $B$, $B_{\text{LSB}}$ | 1. Key-switching. 2. Data hiding by servers | Servers with open access |
| Data hiding key | $k$ | Data hiding and extraction from ciphertext by servers | Servers |

*2) Encryption*

For the pixel pair $(X, Y)$, whose $i$LSBs are denoted by $b_X^i$, $b_Y^i$ ($i=1, 2, \ldots, 8$), each bit is encrypted by LWE encryption with a new public key. We omit the symbol "$_A$" in Eq. (24) for short in this paper: $c_X^i = Enc(b_X^i)$, $c_Y^i = Enc(b_Y^i)$, $i=1, 2, \ldots, 8$.

*3) FHE encapsulated DE data hiding*

  *a) Calculation circuits design*

To realize FHE encapsulated DE, we designed the calculation circuits as Fig. 3 shows. Compared with traditional circuits, we made some simplification on the carry or borrow cases of the highest/lowest bit and the positive or negative sign judgment, which was mainly based on the bit length of pixels and the overflow/underflow constraints in preprocessing. It should be noted that homomorphic circuits share the same internal relationships and operation types as the above calculation circuits, except that it is arithmetic operations which are performed modulo 2 in calculation circuits (Fig. 3(a)) while those would be matrix operations and performed modulo $q$ in homomorphic circuits (Fig. 3(b)).

Pixel adding circuit $Add^*$ of $(X+Y)$ in binary form is as following:

$$(b_{sum}^8, b_{sum}^7, \ldots, b_{sum}^1) = Add^*(b_X^8, b_X^7, \ldots, b_X^1; b_Y^8, b_Y^7, \ldots, b_Y^1) \quad (31)$$

In $Add^*$, there are eight *refreshing* in order from 1 to 8 due to the bit carry case. After each refreshing, one more bit of the sum would be outputted and the inputs would be refreshed. In *refreshing i* ($i=1$ to 8): $b_{sum}^i = b_X^i + b_Y^i$, $b_X^{i+j} = b_X^{i+j} + b_Y^i \cdot (b_X^{i+j-1} b_X^{i+j-2} \ldots b_X^i)$, $b_Y^{i+j} = b_Y^{i+j}$, $j=1,2,\ldots,8-i$.

In this section, assuming $X > Y$, the subtracting circuit $Sub^*$ of $(X-Y)$ is designed (we could confirm the bigger one between a pair $(X, Y)$ according to the map $\boldsymbol{M}_{ava}$):

$$(b_{dif}^8, b_{dif}^7, \ldots, b_{dif}^1) = Sub^*(b_X^8, b_X^7, \ldots, b_X^1; b_Y^8, b_Y^7, \ldots, b_Y^1) \quad (32)$$

In $Sub^*$, there are eight *refreshing* in order from 1 to 8 due to the bit borrow case. The internal relationship between inputs and outputs is exhibited in Fig. 3(c). After each refreshing, one more bit of the difference would be outputted and the minuend would be refreshed. In *refreshing i* ($i=1$ to 8): $b_{dif}^1 = b_X^1 + b_Y^1$, $b_{dif}^i = b_{temp}^i + b_Y^i$, $b_{temp}^{i+j} = b_{temp}^{i+j} + b_Y^i \cdot (b_{temp}^{i+j-1}+1)(b_{temp}^{i+j-2}+1)\ldots(b_{temp}^i+1)$, $b_Y^{i+j} = b_Y^{i+j}$, $j=1,2,\ldots,8-i$.

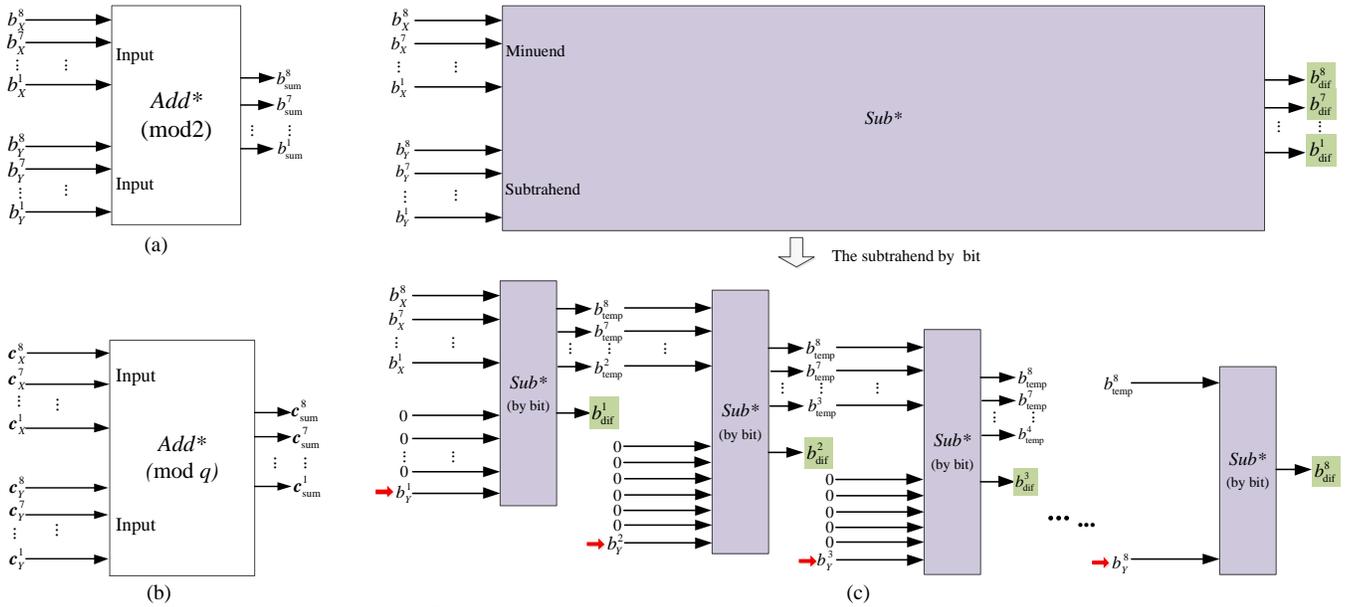

Fig. 3. Circuit sketch: (a) The pixel adding circuit $Add^*$; (b) The homomorphic adding circuit; (c) The pixel subtracting circuit $Sub^*$.

  *b) FHE encapsulated DE data hiding*

Denote the $i$LSB of $h$ and $l$ as $b_h^i$ and $b_l^i$, and the additional bit is $b_s$. $\boldsymbol{c}_h^i = Enc(b_h^i)$, $\boldsymbol{c}_l^i = Enc(b_l^i)$ ($i=1, 2, \ldots, 8$).

Step 1: Calculate $\boldsymbol{c}_h^i$ ($i=1, 2, \ldots, 8$) $\Leftrightarrow (\boldsymbol{c}_h^8, \boldsymbol{c}_h^7, \ldots, \boldsymbol{c}_h^1)$. According to Eq. (1), $(\boldsymbol{c}_h^8, \boldsymbol{c}_h^7, \ldots, \boldsymbol{c}_h^1) = Sub^*(\boldsymbol{c}_X^8, \boldsymbol{c}_X^7, \ldots, \boldsymbol{c}_X^1; \boldsymbol{c}_Y^8, \boldsymbol{c}_Y^7, \ldots, \boldsymbol{c}_Y^1)$.

Step 2: Calculate the encrypted $(X+Y) \Leftrightarrow (\boldsymbol{c}_{sum}^8, \boldsymbol{c}_{sum}^7, \ldots, \boldsymbol{c}_{sum}^1) = Add^*(\boldsymbol{c}_X^8, \boldsymbol{c}_X^7, \ldots, \boldsymbol{c}_X^1; \boldsymbol{c}_Y^8, \boldsymbol{c}_Y^7, \ldots, \boldsymbol{c}_Y^1)$.

Step 3: Calculate $\boldsymbol{c}_{temp0} = Enc(0)$, and the encrypted $l$, $(\boldsymbol{c}_l^8, \boldsymbol{c}_l^7, \ldots, \boldsymbol{c}_l^1)$, can be obtained according to Eq. (2): $(\boldsymbol{c}_l^8, \boldsymbol{c}_l^7, \ldots, \boldsymbol{c}_l^1) = (\boldsymbol{c}_{temp0}, \boldsymbol{c}_{sum}^8, \boldsymbol{c}_{sum}^7, \ldots, \boldsymbol{c}_{sum}^2)$.

Step 4: Calculate $\boldsymbol{c}_{bs} = Enc(b_s)$ and refresh $\boldsymbol{c}_{temp0} = Enc(0)$. The encrypted $h' \Leftrightarrow (\boldsymbol{c}_{h'}^8, \boldsymbol{c}_{h'}^7, \ldots, \boldsymbol{c}_{h'}^1)$ can be obtained according to Eq. (5): $(\boldsymbol{c}_{h'}^8, \boldsymbol{c}_{h'}^7, \ldots, \boldsymbol{c}_{h'}^1) = Add^*(\boldsymbol{c}_h^7, \boldsymbol{c}_h^6, \ldots, \boldsymbol{c}_h^1, \boldsymbol{c}_{temp0}; \boldsymbol{c}_{temp0}, \ldots, \boldsymbol{c}_{temp0}, \boldsymbol{c}_{bs})$.

Step 5: Calculate $\boldsymbol{c}_{temp1} = Enc(1)$. And calculate the encrypted $(h'+1) \Leftrightarrow (\boldsymbol{c}_{sum'}^8, \boldsymbol{c}_{sum'}^7, \ldots, \boldsymbol{c}_{sum'}^1) = Add^*(\boldsymbol{c}_{h'}^8, \boldsymbol{c}_{h'}^7, \ldots, \boldsymbol{c}_{h'}^1; \boldsymbol{c}_{temp0}, \boldsymbol{c}_{temp0}, \ldots, \boldsymbol{c}_{temp0}, \boldsymbol{c}_{temp1})$.

Step 6: Refresh $\boldsymbol{c}_{temp0} = Enc(0)$. The encrypted $X'$ and $Y'$ after DE data hiding are restored according to Eqs. (3)-(4):

$$(\boldsymbol{c}_{X'}^8, \boldsymbol{c}_{X'}^7, \ldots, \boldsymbol{c}_{X'}^1) = Add^*(\boldsymbol{c}_l^8, \boldsymbol{c}_l^7, \ldots, \boldsymbol{c}_l^1; \boldsymbol{c}_{temp0}, \boldsymbol{c}_{sum'}^8, \boldsymbol{c}_{sum'}^7, \ldots, \boldsymbol{c}_{sum'}^2) \quad (33)$$

$$(\boldsymbol{c}_{Y'}^8, \boldsymbol{c}_{Y'}^7, \ldots, \boldsymbol{c}_{Y'}^1) = Sub^*(\boldsymbol{c}_l^8, \boldsymbol{c}_l^7, \ldots, \boldsymbol{c}_l^1; \boldsymbol{c}_{temp0}, \boldsymbol{c}_{h'}^8, \boldsymbol{c}_{h'}^7, \ldots, \boldsymbol{c}_{h'}^2) \quad (34)$$

Following each homomorphic multiplication, key-switching is implemented to eliminate the secondary data expansion of ciphertext. The bootstrapping is implemented for every 10 homomorphic multiplication or 100 homomorphic addition to control noise excessive stacking.

### 4) Key-switching based LSB data hiding

Step 1: Randomly scramble the additional data sequence $b_s$ by using data hiding key $k$ to obtain the to-be-embedded data $b_r$:

$$b_r = k \oplus b_s \quad (35)$$

where $b_r \in b_r$. Denote the last element of $c_X^1$ as $c_{LX1}$, whose LSB would be replaced by $b_r$ ($X$ is the "1" signed pixel by $M_{ava}$).

Step 2: If $b_r = LSB(c_{LX1})$, $c_X^1$ maintains the same, or if $b_r \neq LSB(c_{LX1})$, $c_X^1$ is refreshed by: $c_X^1 = BitDe(c_X^1)^T B_{LSB}$.

Step 3: Repeat Step 2 until $LSB(c_{LX1}) = b_r$.

The marked ciphertext is obtained: $c_X^i$ and $c_Y^i$ ($i=1, 2, \ldots, 8$).

According to the framework in Fig. 2(a), after receiving the marked ciphertext, the client user could implement the decryption on the marked ciphertext to obtain $X'$ and $Y'$ by using $s$: $b_{X'}^i = Dec_s(c_{X'}^i)$, $b_{Y'}^i = Dec_s(c_{Y'}^i)$, ($i=1, 2, \ldots, 8$). The additional data could be extracted according to DE extraction (Eq. (6)) and the pixels could be recovered according to DE recovery (Eqs. (7), (3), and (4)).

### 5) LSB extraction from the marked ciphertext

Additional data could be directly extracted from ciphertext without the private key $s$ ($X$ is the "1" signed pixel by $M_{ava}$):

$$b_r = LSB(c_{LX1}) \quad (36)$$
$$b_s = k \oplus b_r \quad (37)$$

### 6) FHE encapsulated DE recovery

FHE encapsulated DE recovery is implemented by the servers to return a new ciphertext corresponding to the plaintext without additional data embedded.

Step 1: Calculate $c_{h'}^i$ ($i=1, 2, \ldots, 8$) by using ($c_{X'}^8, c_{X'}^7, \ldots, c_{X'}^1$)($c_{Y'}^8, c_{Y'}^7, \ldots, c_{Y'}^1$) according to Eq. (1): ($c_{h'}^8, c_{h'}^7, \ldots, c_{h'}^1$) = $Sub^*(c_{X'}^8, c_{X'}^7, \ldots, c_{X'}^1; c_{Y'}^8, c_{Y'}^7, \ldots, c_{Y'}^1)$.

Step 2: Calculate $c_{temp0} = Enc(0)$. The encrypted $h \Leftrightarrow (c_h^8, c_h^7, \ldots, c_h^1)$ can be obtained according to Eq. (7): $(c_h^8, c_h^7, \ldots, c_h^1) = (c_{temp0}, c_{h'}^8, c_{h'}^7, \ldots, c_{h'}^2)$.

Step 3: Calculate the encrypted $(X'+Y') \Leftrightarrow (c_{sum''}^8, c_{sum''}^7, \ldots, c_{sum''}^1) = Add^*(c_{X'}^8, c_{X'}^7, \ldots, c_{X'}^1; c_{Y'}^8, c_{Y'}^7, \ldots, c_{Y'}^1)$.

Step 4: Refresh $c_{temp0} = Enc(0)$, and the encrypted $l$, ($c_l^8, c_l^7, \ldots, c_l^1$), can be obtained according to Eq. (2): ($c_{temp0}, c_{sum''}^8, c_{sum''}^7, \ldots, c_{sum''}^2$).

Step 5: Calculate $c_{temp1} = Enc(1)$. And calculate the encrypted $(h+1) \Leftrightarrow (c_{sum'''}^8, c_{sum'''}^7, \ldots, c_{sum'''}^1) = Add^*(c_h^8, c_h^7, \ldots, c_h^1; c_{temp0}, c_{temp0}, \ldots, c_{temp0}, c_{temp1})$.

Step 6: Refresh $c_{temp0} = Enc(0)$. The encrypted $X$ and $Y$ are restored according to Eqs. (3)-(4):

$$(c_X^8{}', c_X^7{}', \ldots, c_X^1{}') = Add^*(c_l^8, c_l^7, \ldots, c_l^1;$$
$$c_{temp0}, c_{sum'''}^8, c_{sum'''}^7, \ldots, c_{sum'''}^2) \quad (38)$$
$$(c_Y^8{}', c_Y^7{}', \ldots, c_Y^1{}') = Sub^*(c_l^8, c_l^7, \ldots, c_l^1;$$
$$c_{temp0}, c_h^8, c_h^7, \ldots, c_h^2) \quad (39)$$

According to the Fig. 2(a), after receiving the restored ciphertext, the client user could implement the decryption to obtain the original pixels $X$ and $Y$ by using $s$: $b_X^i = Dec_s(c_X^i{}')$, $b_Y^i = Dec_s(c_Y^i{}')$, ($i=1, 2, \ldots, 8$).

### 7) FHE encapsulated DE extraction

It shares the same step as Step 1 in *FHE encapsulated DE recovery* to obtain the encrypted $h' \Leftrightarrow (c_{h'}^8, c_{h'}^7, \ldots, c_{h'}^1)$. The encrypted $b_s$ is $c_{h'}^1$.

After receiving the encrypted additional data, the client user could implement the decryption to obtain the embedded data: $b_s = Dec_s(c_{h'}^1)$.

## C. Efficient FHEE-DE

### 1) Preprocessing of DE

Efficient FHEE-DE shares the same preprocessing as universal FHEE-DE.

### 2) Encryption

The client user calculates the $(h, l)$ of $(X, Y)$ first. The $(h, l)$ would be encrypted as ciphertext which would be uploaded to the server: $c_h^i = Enc(b_h^i)$ and $c_l^i = Enc(b_l^i)$ ($i=1, 2, \ldots, 8$).

### 3) FHE encapsulated DE data hiding

Calculate $c_{bs} = Enc(b_s)$ and $c_{temp0} = Enc(0)$. The encrypted $h' \Leftrightarrow (c_{h'}^8, c_{h'}^7, \ldots, c_{h'}^1)$ can be obtained according to Eq. (5): $(c_{h'}^8, c_{h'}^7, \ldots, c_{h'}^1) = Add^*(c_h^7, c_h^6, \ldots, c_h^1, c_{temp0}; c_{temp0}, c_{temp0}, \ldots, c_{temp0}, c_{bs})$.

Return $c_{h'}^i$ and $c_l^i$ ($i=1, 2, \ldots, 8$) as the DE embedded ciphertext.

### 4) Key-switching based LSB data hiding

Step 1: The same as Step1 of *Key-switching based LSB data hiding* in universal FHEE-DE. Denote the last element of $c_{h'}^1$ as $c_{Lh1}$, whose LSB would be replaced by $b_r$.

Step 2: If $b_r = LSB(c_{Lh1})$, $c_{h'}^1$ maintains the same, or if $b_r \neq LSB(c_{Lh1})$, $c_{h'}^1 = BitDe(c_{h'}^1)^T \cdot B_{LSB}$.

Step 3: Repeat Step 2 until $LSB(c_{Lh1}) = b_r$.

The marked ciphertext is obtained: $c_{h'}^i$ and $c_l^i$ ($i=1, 2, \ldots, 8$).

After receiving the marked ciphertext, the client user could decrypt it and obtain the marked $(h', l')$ by using $s$: $b_{h'}^i = Dec_s(c_{h'}^i)$, $b_l^i = Dec_s(c_l^i)$, ($i=1, 2, \ldots, 8$). Then DE extraction and recovery could be implemented.

Efficient FHEE-DE shares the same *LSB extraction from the marked ciphertext* as universal FHEE-DE.

### 5) FHE encapsulated DE recovery

Calculate $c_{temp0} = Enc(0)$. The encrypted $h \Leftrightarrow (c_h^8, c_h^7, \ldots, c_h^1)$ can be obtained according to Eq. (7): $(c_h^8, c_h^7, \ldots, c_h^1) = (c_{temp0}, c_{h'}^8, c_{h'}^7, \ldots, c_{h'}^2)$. The unmarked $c_h^i$ and $c_l^i$ ($i=1, 2, \ldots, 8$) are obtained. The client user could decrypt it to obtain $h$ and $l$, and restore $X$ and $Y$ losslessly.

### 6) FHE encapsulated DE extraction

The encrypted $b_s$ is obtained by $c_{bs} = c_{h'}^1$. The client user could decrypt it to obtain the embedded data.

## V. THEORETICAL ANALYSIS AND EXPERIMENTAL RESULTS

### A. Correctness

The correctness of the proposed scheme includes the lossless restoration of plaintext and the accurate extraction of the embedded data. The test images, 512×512 8-bit grayscale images, are from image libraries, USC-SIPI

(http://sipi.usc.edu/database/database.php?volume= misc) and Kodak (http://r0k.us/graphics/kodak/index.html). The experimental results of six test images were selected in this section to demonstrate the correctness. The six test images are as shown in Fig. 4. The preprocessing of DE, LWE encryption & decryption, key switching, and KS-LSB were all implemented on MATLAB2010b with a 64-bit single core (i7-6800K) @ 3.40GHz. We referred to the method in [36] to realize bootstrapping, of which implementation tools and source codes are available on https://github.com/tfhe/tfhe, implemented in *C++* on a NVIDIA Titan-XP GPU card. We performed bootstrapping operations after each homomorphic circuit calculation.

*Parameters setting*: Solving the LWE problem with given parameters is equivalent to solving Shortest Vector Problem (SVP) in a lattice with a dimension $\sqrt{n\log_2(q)/\log_2(\delta)}$. Considering the efficiencies of the best known lattice reduction algorithms, the secure dimension of the lattice must reach 500 ($\delta=1.01$) [37], [38]. An increase in $n$ will result in a high encryption blowup. To balance security and the efficiency of practical use, we set $n=240$, $q=57601$, $d=4573$. To ensure the fidelity of the marked plaintext, we set $h_{\mathrm{fid}}=10$.

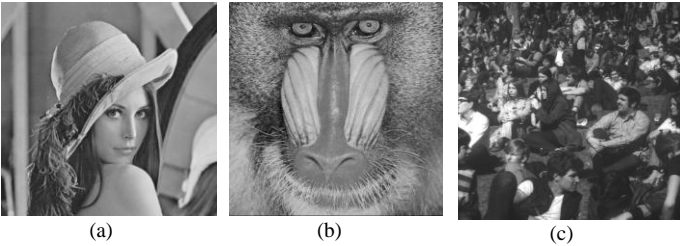
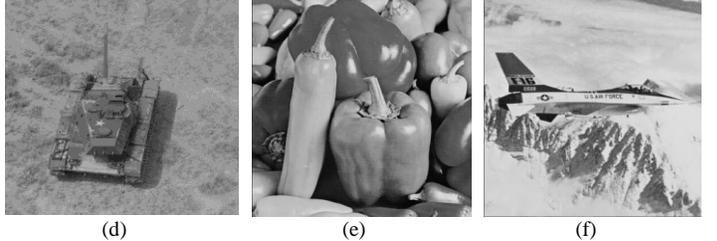

(a)    (b)    (c)    (d)    (e)    (f)

Fig. 4. The test images. (a) Lena; (b) Baboon; (c) Crowd; (d) Tank; (e) Peppers; (f) Plane.

*1) Reversibility of plaintext recovery*

In the proposed scheme, there are two cases for plaintext recovery: *a)* the user directly decrypts the marked ciphertext to get the marked plaintext. We calculated the PSNR of the marked plaintext, named by PSNR1. And then the plaintext can be obtained after DE recovery. We then calculated the PSNR of the recovered plaintext, named by PSNR2. *b)* The third server implements FHEE-DE recovery on the marked ciphertext to obtain new ciphertext. The user receives the new ciphertext and decrypts it to obtain the plaintext. The PSNR of the plaintext is named by PSNR3.

In the experiment, we obtained the available pixel pairs for difference extension according to the constraints in Eqs. (18)-(20). For the ciphertext of the available pixel pairs, two encrypted pixels would carry one bit of additional data by FHEE-DE, and an extra bit by KS-LSB. However, the actual content of the two bits of embedded data is the same. We counted the two bits of embedded data as one bit of embedding capacity. Therefore, the maximum EC of an image is only related to the number of the available pixel pairs.

The values of PSNR of DE in [15] and PSNR1-3 with the maximum EC are listed in Table II. From the results of PSNR1, it could be seen that there is embedding distortion in the marked plaintext of FHEE-DE, and the degree of distortion is lower than DE in [15]. Both PSNR2 and PSNR3 are "∞", indicating that the recovered plaintext has achieved no distortion.

We continue to analyze the PSNR1 of the test images at different EC. Fig. 5 shows that the principle of DE is to improve the fault-tolerance ability of the difference between pixels through difference expansion, so that additional information can be reversibly loaded. Therefore, the smaller the difference is, the smaller modification the embedded pixel has. When performing a non-full-embedded experiment, the available pixel pairs were preferentially selected with a smaller $h_{\mathrm{fid}}$.

TABLE II
THE PSNR (*dB*) OF ED AND FHEE-DE WITH THE MAXIMUM EC(*bits*).

| Image | Max EC | PSNR in [15] | PSNR1 | PSNR2 | PSNR3 |
|---|---|---|---|---|---|
| Lena | 110195 | 33.6360 | 42.1171 | ∞ | ∞ |
| Baboon | 69286 | 29.2830 | 41.3894 | ∞ | ∞ |
| Crowd | 104882 | 33.2144 | 42.4764 | ∞ | ∞ |
| Tank | 108963 | 35.3958 | 40.4472 | ∞ | ∞ |
| Peppers | 110558 | 34.0715 | 40.5025 | ∞ | ∞ |
| Plane | 114834 | 33.4095 | 42.8519 | ∞ | ∞ |
| Average | 103120 | 33.1684 | 41.6308 | ∞ | ∞ |

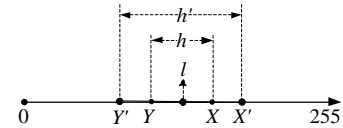

Fig. 5. The principle of difference expansion.

The marked images of Lena with different $h_{\mathrm{fid}}$ are as shown in Fig. 6. In Table III, we list the EC at different $h_{\mathrm{fid}}$ and the corresponding PSNR1 values of the six test images. Then a comparison of the performance of PSNR1 was made between the three RDH-ED methods [23], [25], [26] and the proposed one. The results of test images from USC-SIPI and Kodak show that the proposed scheme has a better fidelity of the mark plaintext directly decrypted from the marked ciphertext. As shown in Fig. 7, we demonstrated the PSNR1 of the methods in [23], [25], [26] and the proposed scheme on the test images Lena and Plane.

Compared with RDH-ED, methods of RDH for the spatial domain have better fidelity of the marked plaintext and more technical implementations. The introduction of homomorphic techniques into RDH-ED can provide a technical bridge for introducing more existing RDH methods into the encrypted domain.

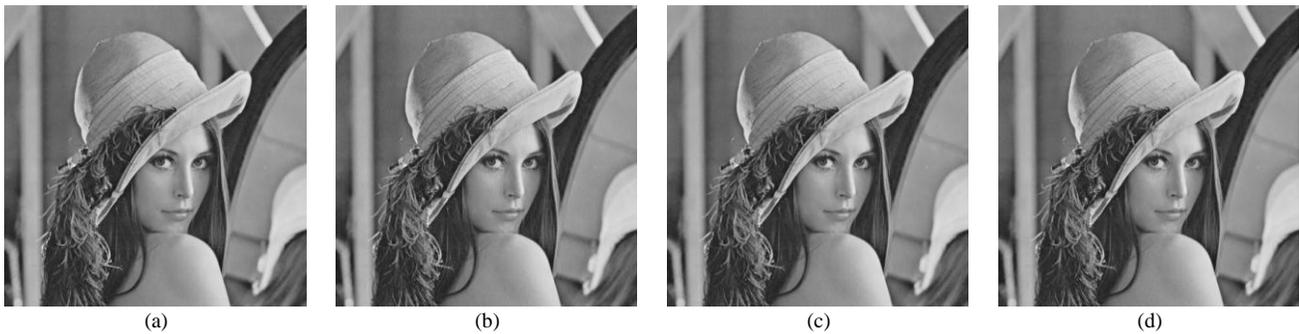

(a)            (b)            (c)            (d)

Fig. 6. The test images. (a) Original Lena; (b) Marked Lena with $h_{fid}$=10, EC=110195 bits, PSNR1= 42.1171 dB; (c) Marked Lena with $h_{fid}$=2, EC=50232 bits, PSNR1= 52.2932 dB; (d) Marked Lena with $h_{fid}$=0, EC=11434 bits, PSNR1= 64.7369 dB.

TABLE III
PSNR1 (*dB*) VERSUS EC(*bits*) AT DIFFERENT $h_{fid}$.

| Image | $h_{fid}$=5 | | $h_{fid}$=3 | | $h_{fid}$=2 | | $h_{fid}$=1 | | $h_{fid}$=0 | |
|---|---|---|---|---|---|---|---|---|---|---|
|  | EC | PSNR1 | EC | PSNR1 | EC | PSNR1 | EC | PSNR1 | EC | PSNR1 |
| Lena | 86605 | 45.6073 | 65303 | 49.3394 | 50232 | 52.2932 | 32104 | 56.6637 | 11434 | 64.7369 |
| Baboon | 42522 | 47.9412 | 28553 | 52.5559 | 20702 | 55.9498 | 12464 | 60.7332 | 4210 | 69.0304 |
| Crowd | 86962 | 47.0505 | 73240 | 50.1791 | 64164 | 52.1256 | 46810 | 55.9208 | 26504 | 61.0445 |
| Plane | 100746 | 46.1582 | 85200 | 48.9055 | 71114 | 51.2680 | 50764 | 54.8316 | 19966 | 62.3590 |
| Peppers | 80017 | 45.4989 | 56523 | 49.7460 | 42091 | 52.9404 | 26044 | 57.5117 | 8791 | 65.9745 |
| Tank | 77887 | 45.6201 | 53520 | 50.3489 | 43832 | 52.5652 | 20988 | 60.6655 | 16843 | 63.1004 |

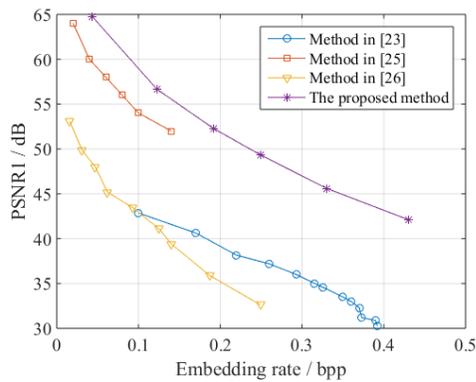

(a)

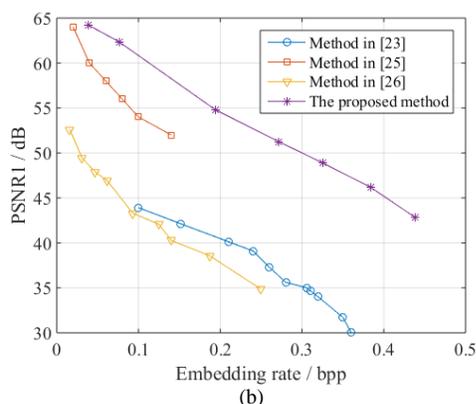

(b)

Fig. 7. PSNR1 (dB) of FHEE-DE with different ER (bpp) on (a) Lena; (b) Plane.

*2) Accuracy of data extraction*

There are three cases of data extraction in this paper. The realization of the three cases is the embodiment of the separability of the proposed scheme:

*a)* The third-party server directly extracts the embedded data from the marked ciphertext by using KS-LSB extraction. Fig. 8(a) shows the comparison result bit by bit between the extracted data and the additional data with an EC of 100000 bits in the experiment. It demonstrates that the extraction accuracy was 100%. *b)* The user decrypts the marked ciphertext to obtain the marked plaintext, and then uses DE extraction to extract data. As shown in Fig. 8(b), the extraction accuracy is 100%. *c)* The third-party server first performs FHEE-DE extraction on the marked ciphertext to obtain the encrypted embedded data. The user decrypts it to obtain the embedded data. The accuracy is as shown in Fig. 8(c).

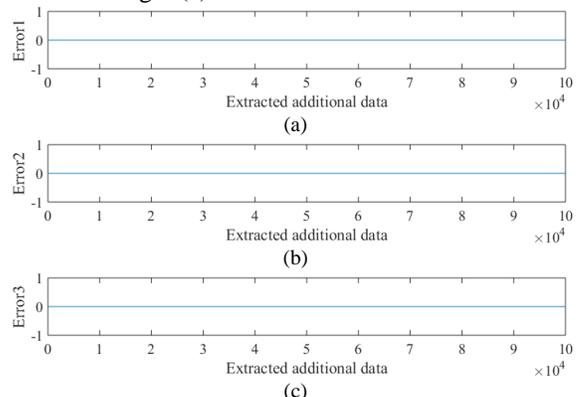

Fig. 8. Errors of the extracted data: (a) Error1 from KS-LSB extraction on the marked ciphertext; (b) Error2 from DE extraction on the marked plaintext; (c) Error3 from LWE decryption on the encrypted additional data.

*B. Security*

Security of RDH-ED mainly includes two aspects: *a)* Data hiding should not weaken the security of the original encryption or leave any hidden danger of security cracking. *b)* The embedded information cannot be obtained by an attacker without the extraction key or the private key.

In [27]-[28], through the derivation of the probability distribution function (PDF) on the marked ciphertext and the experimental analysis of the statistical features, it was proved that the ciphertext distribution before and after data hiding did not change, so that the security of the RDH-ED method was proved by certain reasoning. In this paper, we do not have to make relevant derivation on PDF or the statistical features, because all the homomorphic operations of the proposed scheme are equivalent to the operations of re-encryption [33], and the encryption security can be directly guaranteed by the principles of FHE.

The processes of implementing FHEE-ED and KS-LSB on the ciphertext are equivalent to the processes of re-encrypting the ciphertext, which would not reveal anything about the private key or reduce the encryption security. The additional data is encrypted by LWE encryption before FHEE-DE data hiding, or scrambled using sequence encryption by the third party before KS-LSB data hiding, which ensure the confidentiality of the additional data. During the transmission or processing by third-party servers, the third party does not obtain any information related to the client user's private key, nor did it expose any relationship between plaintext and its corresponding ciphertext. Even if public keys used in the re-encryption are all generated by the same private key, there is a random variable participating in the generation process, *i.e.*, $e$ in Eq. (23), thus ensuring the independence among public keys. Due to the random variable $a_r$ in Eq. (25), different ciphertext encrypted by the same public key would be also independent from each other, even if the different ciphertext were corresponding to the same plaintext. That is the advantage of the public key cryptosystem in application. In summary, the security of the proposed scheme can realize the security that LWE encryption has achieved. What is more, the security of LWE encryption reaches anti-quantum algorithm analysis, while Paillier algorithms cannot resist quantum algorithm analysis.

*C. Efficiency*

*1) Public key consumption*

As shown in Table I, there are three types of public keys: $A$, $B$, and $B_{LSB}$. Although they are not stored locally after generation, their consumption is directly related to the number of the operations of key switching and bootstrapping during data hiding, which determines the efficiency of FHEE-DE. Therefore, it is necessary to analyze the public key consumption in different types of operations in FHEE-DE. The analysis of addition circuit (*Add\**) is as follows:

In *refreshing i* ($i$=1,2,…,8): There are 1 bit-addition and $\sum_{\mu=1}^{i-1}\mu$ bit-multiplication. Therefore, the total amount of homomorphic addition is 8, and the total amount of homomorphic multiplication is 84. In the same way, we got Table IV.

TABLE IV
THE NUMBERS OF TYPES OF OPERATIONS

| Circuit | Reflesh $i$ ($i$=1-8) | | Total amount | | | | |
|---|---|---|---|---|---|---|---|
| | + | × | + | × | Key-switching | Boot-strapping | Public key |
| *Add\** | 1 | $\sum_{\mu=1}^{8-i}\mu$ | 8 | 84 | 84 | 9 | 93 |
| *Sub\** | $\sum_{\mu=1}^{9-i}\mu$ | $\sum_{\mu=1}^{8-i}\mu$ | 120 | 84 | 84 | 9 | 93 |

In universal FHEE-DE, there are 7 *add\** and 3 *sub\**. In efficient FHEE-DE, there are 1 *add\** and 0 *sub\**. Obviously, compared with the universal FHEE-DE, the key consumption is reduced from $o(100)$ to $o(10)$ in efficient FHEE-DE. Therefore, the number of key switching and bootstrapping can be also reduced greatly. Since the computational complexity and the elapsed time of bootstrapping are much higher than other processes, i.e., encryption, decryption, or key switching, the efficient FHEE-DE has higher operational efficiency than universal FHEE-DE.

The secondary expansion of ciphertext needs to be eliminated by using key switching technique. A new ciphertext can be obtained by performing only once matrix multiplication between a switching key and the old ciphertext, which is fast and can ensure the confidentiality of plaintext and the private key.

KS-LSB data hiding is to randomly change the LSB of specific ciphertext by key switching until the LSB is the same as the to-be-embedded bit. Let the number of times of key switching performed for one bit embedding be $\lambda$, that is, the public key consumption of KS-LSB for one bit embedding is $\lambda$. Since the LSB of the ciphertext is 0 or 1 randomly appeared with a probability of 0.5, $\lambda$+1 obeys the geometric distribution as shown in Table V. It demonstrates that it would be a small probability event with a probability less than 3% to operate more than 4 times key switching to realize one bit embedding. The theoretical value of $\lambda$ is 0.8906. In the experiment, we performed 1000 KS-LSB data hiding tests. The actual $\lambda$ was 0.995 on average, indicating a high embedding accuracy and efficiency.

TABLE V
THE PROBABILITY DISTRIBUTION OF $\beta$

| $\lambda$ | 0 | 1 | 2 | 3 | 4 | 5 |
|---|---|---|---|---|---|---|
| $P$ | 0.5 | 0.25 | 0.125 | 0.0625 | 0.0313 | 0.0156 |

*2) Elapsed time*

The public key encryption algorithms, including the Paillier algorithm and the LWE algorithm, have ciphertext extension. In [28], the ciphertext extension of Paillier and LWE encryption was discussed in detail. Due to the application of the separability of RDH-ED, ciphertext is usually stored in the server or the cloud, the local storage cost of users is not too much. However, the elapsed time of encryption, decryption, data hiding, and data extraction is related to the efficiency in practice. In this section, we mainly demonstrate the elapsed time of each operation. Table VI lists the elapsed time of the four main operations in FHEE-DE (the elapsed time of bootstrapping was obtained on the GPU card, and the others were obtained on the CPU card).

The elapsed time is specifically the time (milliseconds) when one bit plaintext gets decrypted, or one public key is generated and consumed by the operation, e.g., one bit of plaintext gets encrypted after each elapsed time of encryption.

TABLE VI
ELAPSED TIME (*ms*) OF ONCE OPERATION IN FHEE-DE

| Operation | Encryption | Decryption | Key switching | Bootstrapping |
|---|---|---|---|---|
| Elapsed time | 20.5971 | 0.0067 | 0.1054 | *0.4922* |

The brief structure and linear operations of LWE provide LWE based FHEE-DE with a low time consumption, which are significant in practice. The results in Table VI indicate that the elapsed time of the proposed method is acceptable for practical use.

## VI. Conclusion

The main contributions of this paper are as following:

1. We propose a novel scheme of FHEE-DE to realize data hiding in encrypted domain based on LWE. A fidelity constraint is proposed, which has enhanced the PSNR1 of the marked plaintext. Techniques of key-switching and bootstrapping are introduced to control the ciphertext extension and decryption failure, which might contribute to introduce other existing RDH methods.

2. KS-LSB data hiding in the encrypted domain has been proposed, which supports the extraction directly from the encrypted domain without the private key.

3. We propose an efficient version of FHEE-DE, which could remarkably improve the efficiency compared with the universal FHEE-DE by simplifying FHE operations. However, the technical scalability is not as good as the universal FHEE-DE.

Experimental results demonstrate that the performances of the proposed scheme in EC, fidelity, and reversibility are superior to most existing RDH-ED methods, and fully separability was achieved without reducing the security of LWE encryption. Future investigation will focus on introducing more RDH methods in image spatial domain into the encrypted domain and optimizing the technique of FHEE-DE to further improve the efficiency.

## VII. Acknowledgements

This work was supported by National Key Research & Development Program of China under Grant No. 2017YFB0802000, and the National Natural Science Foundation of China under Grant No.61379152, Grant No. 61872384 and Grant No.61403417.

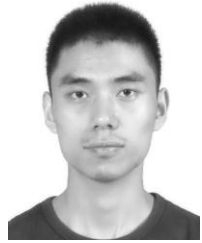

**Yan Ke** received the M.S. degree in cryptography from Engineering University of PAP, Xi'an, China in 2016. He is currently pursuing the Ph.D. degree in cryptography at Engineering University of PAP. His research interest includes reversible data hiding, lattice cryptography.

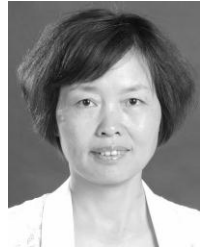

**Minqing Zhang** received the M.S. degree in computer science & application from Northwestern Polytechnical University, Xi'an, China, in 2001, and Ph.D. degree in network & information security from Northwestern Polytechnical University, Xi'an, China, in 2016. Currently, she has been with the Key Laboratory of Network and Information Security under Chinese People Armed Police Force as a professor. Her research interests include cryptography, and trusted computation.

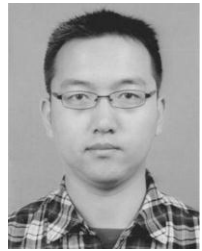

**Jia Liu** received the M.S. degree in cryptography from Engineering University of PAP, Xi'an, China, in 2007 and Ph.D. degree in neural network and machine learning from Shanghai Jiao Tong University, Shanghai, China, in 2012. Currently, he has been with the Key Laboratory of Network and Information Security under PAP as an associate professor. His research interests include pattern recognition and image processing.

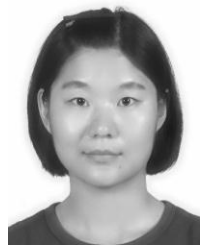

**Tingting Su** received the B.S. degree in information research & security from Engineering University of PAP, Xi'an, China in 2010 and the M.S. degree in cryptography from Engineering University of PAP, Xi'an, China in 2013. Her research interest includes mathematics statistics and cryptography.

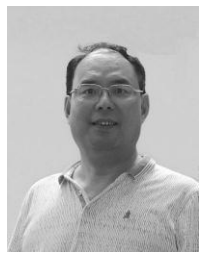

**Xiaoyuan Yang** received the B.S. degree in applied mathematics from Xidian University, Xi'an, China, in 1982 and the M.S. degree in cryptography & encoding theory from Xidian University, Xi'an, China, in 1991. Currently, he has been with the Key Laboratory of Network and Information Security under PAP as a professor. His research interests include cryptography and trusted computation.